\documentclass[aps,prd,twocolumn,showpacs]{revtex4-2}
\usepackage{graphicx}
\usepackage{amsmath}
\usepackage{amssymb}
\usepackage{sistyle}
\usepackage{color}

\begin{document}

\title{Zigzag edge states in graphene in
the presence of in-plane electric field}

\author{A.A.~Herasymchuk}
%\thanks{On leave of absence from }
%\email{viacheslav.tsaran@gmail.com}
\affiliation{Department of Physics, Taras Shevchenko National University of Kyiv,
64/13, Volodymyrska Street, Kyiv 01601, Ukraine}

\author{S.G.~Sharapov}
%\email{sharapov@bitp.kiev.ua}
\affiliation{Bogolyubov Institute for Theoretical Physics, National Academy of Science of Ukraine, 14-b Metrologichna Street, Kyiv, 03143, Ukraine}
\affiliation{
Kyiv Academic University, 03142 Kyiv, Ukraine}

\author{V.P.~Gusynin}
%\email{sharapov@bitp.kiev.ua}
\affiliation{Bogolyubov Institute for Theoretical Physics, National Academy of Science of Ukraine, 14-b Metrologichna Street, Kyiv, 03143, Ukraine}

\date{\today }

\begin{abstract}
The present study explores the edge states in a finite-width graphene ribbon and a semi-infinite geometry subject to a perpendicular magnetic field and an in-plane electric field, applied perpendicular to a zigzag edge.
To accomplish this, a combination of analytic and numerical methods within the framework of low-energy effective theory  is employed.
Both the gapless and gapped Dirac fermions in graphene are considered.
It is found that a surface mode localized at the zigzag edge remains
dispersionless even in the presence of electric field. This
is shown analytically by employing Darwin's expansion of
the parabolic cylinder functions of large order and argument.
\end{abstract}

%\pacs{71.70.Di, 72.80.Vp}
%Landau levels, 71.70.Di

%\keywords{graphene ribbon, zigzag edge, dispersionless edge mode, Landau level collapse}

\maketitle

\section{Introduction}

Graphene continues to be a playground for investigating
a plethora of unusual electronic phenomena, making it a fascinating system
for fundamental studies in condensed matter physics.
The fact that graphene is  two-dimensional
material allows to access the regime when the confining
potential at the edges of graphene nanoribbons is
atomically sharp.
The quantum Hall edge states in this case are defined by
boundary conditions of vanishing electron wave functions at the
crystal edges.

We recall that in the conventional experiments on two-dimensional
semiconductors only %the opposite limit with
the regime with electrostatically reconstructed edges is accessible.
In this case the system lowers
its energy by reconstructing the edge states into steps which
produce alternating compressible and incompressible
stripes \cite{Chklovskii1992PRB}. Furthermore, imaging these
edge states is difficult because they are buried inside the
semiconductors.

Therefore graphene provides an opportunity to explore the real-space
structure of the edge states by scanning probe techniques
\cite{Li2012NatCom,Coissard2022} avoiding their electrostatic
reconstruction.
Other techniques of vizualization of charge
transport through Landau levels are also available. For example,
one can rely on the scanning photocurrent microscopy \cite{Nazin2010NatPhys}
to use an engineered array of near-surface, atomic-sized quantum sensors \cite{Tetienne2017ASc} or apply high-resolution atomic force microscopy
\cite{Kim2021NarCom}.

An additional opportunity provided by graphene is a mechanism
for manipulating the transport channels by using an external
in-plane Hall electric field. Theoretical exploration of this mechanism
is based on description of graphene layer in
crossed uniform electric and magnetic fields.

The behavior of electrons in crossed electric and magnetic fields
was studied by considering the $3+1$-dimensional Dirac equation.
For example, it was shown  that there are no relativistic corrections
to the quantized Hall
conductivity \cite{MacDonald1983PRB,Nieto1985AJP}.

After discovery of graphene an attention to $2+1$-relativistic fermions
in crossed fields was brought because of the spectacular phenomenon of Landau level collapse. As the dimensionless parameter $\beta = c E/(v_F H)$
reaches its critical value, $|\beta_c | =1$,  the Landau level staircase merges \cite{Lukose2007PRL,Peres2007JPCM}. Here $H$ is a magnetic field $H$ applied perpendicular to the sheet of graphene, $E$ is an applied in-plane
electric field $E$, $v_F$ is the Fermi velocity and CGS units
are used.
The observation of the Landau level collapse was reported in Refs.~\cite{Singh2009PRB,Gu2011PRL}.

The spectrum of an infinite graphene's sheet in presence
of crossed uniform electric and magnetic
fields was investigated analytically in Ref.~\cite{Lukose2007PRL}
by means of a ``Lorenz boost'' transformation that eliminates
the electric field and thus reduces the problem of finding
spectrum to the known one.
The same problem was addressed in Ref.~\cite{Peres2007JPCM} using
algebraic methods.
The influence of a Hall electric field on the Hall
conductivity in graphene was analytically studied in
Ref.~\cite{Krstajic2011PRB} using the spectrum and wave functions
found in Refs.~\cite{Lukose2007PRL,Peres2007JPCM}.
Another possibility to realize the Landau-level collapse would be by generating strain induced either pseudomagnetic or electric fields
suggested in Refs.~\cite{Castro2017PRB,Grassano2020PRB}, respectively
(see also, for example, Refs.~\cite{Castro-Villarreal2017PRB,Morales2023PRB}
on a recent progress in curved spacetime Dirac equation approach
for generation of these fields).
The creation of pseudomagnetic field in graphene was
proven experimentally in Ref. \cite{Levy2010Science}.

As the parameter $\beta$ approaches the critical value, the large values of the involved wave vectors go beyond the range of applicability of long wave length approximation.  Nevertheless, its validity was verified in Ref.~\cite{Lukose2007PRL}
by performing numerical computations on tight binding model
for graphene lattices of a finite size with the
current parallel to the zigzag edges. Still the Landau level collapse
occurs in this case, yet at the lower value of $\beta_c \simeq 0.9$.

Furthermore, numerical computations on the finite lattice in
Ref.~\cite{Lukose2007PRL} [see also Refs.~\cite{Roslyak2010PTRS,Ostahie2015PRB}]
allow to explore the evolution of the edge states in the presence of electric
field. One of the observations is that because of the tilt of the
spectrum, the number of edge channels on the two sides of the zigzag
ribbon becomes different, and current carrying channels appear
in the middle of the stripe \cite{Ostahie2015PRB}.

The ribbons with current along the armchair edges were
considered in  Ref.~\cite{Roslyak2010PTRS}. It is pointed out
that the combined effect of both magnetic and electric fields
breaks the symmetry between electronic and hole energy bands
which is present when either magnetic or electric field is applied.
This paves a way for creation valley filtering devices tunable
by an in-plane electric field \cite{Chen2022} (see also, e.g. Ref.~\cite{Beenakker2007NP}). Various ways of controlling
the electron propagation in graphene nanostructures
with magnetic and electric fields are discussed, for example,
in Refs.~\cite{Tan2009PRB,Milicevic2019PRB,Somrooba2021PE,Fernandes2023PRB}.

In addition to the analytical studies of Landau levels in crossed fields \cite{Lukose2007PRL,Peres2007JPCM}, Landau levels for ribbons and in
semi-infinite geometries were also investigated using the low-energy
model without electric field.
The corresponding differential equation for the spectrum with the appropriate
boundary conditions is treated analytically and then
the eigenenergies are computed numerically by solving
the transcendental equation for gapless \cite{Brey2006bPRB,Ababnin2006PRL,Romanosvsky2011PRB,Wang2011EPJB}
and gapped
\cite{Gusynin2008PRB,Gusynin2008FNT,Gusynin2009PRB} graphene.
A quantitative analytical description of the edge states within WKB
approximation is developed in Ref.~\cite{Deplace2010PRB}
basing on an effective Hamiltonian with a potential depending
on the boundary conditions \cite{Brey2006bPRB,Ababnin2006PRL}.

On the other hand, the consideration of Landau levels in the
crossed fields on the ribbons is limited to the numerical
consideration of the lattice model \cite{Lukose2007PRL,Roslyak2010PTRS,Ostahie2015PRB}.
The purpose of the present work is to extend the
analytical study of graphene nanoribbons in the crossed magnetic
and  in-plane  electric Hall field  applied perpendicular to
the ribbon edges.

The paper is organized as follows.
In Sec.~\ref{sec:model}, we introduce a low-energy model for a graphene ribbon
with a zigzag edge subject to crossed magnetic and
electric fields. A general solution
of the differential equations describing graphene
in terms of the parabolic cylinder Weber functions
is obtained in Sec.~\ref{sec:general-sol} and the spectrum of an infinite
graphene's sheet is reproduced. In Sec.~\ref{sec:ribbom-zigzag},
we present the spectra for a ribbon solving numerically the equations
derived in  Appendix~\ref{sec:Appendix-sol}.
In Sec.~\ref{sec:zigzag-semi}, we present the spectra in a
semi-infinite geometry and obtain the asymptotic
solutions for these spectra for zero electric field
in the bulk and near the edge.
We also present an analytic consideration of the
dispersionless mode specific for the zigzag edges.
The details are provided in Appendix~\ref{sec:Appendix-dispersionless}.
In the Conclusion (Sec.~\ref{sec:conclusion}), we summarize the
obtained results and discuss
their possible experimental observation.

\section{Model}
\label{sec:model}

To determine eigenenergy $\mathcal{E}$
we consider the stationary Dirac equation
\begin{equation}
\label{Dirac-E}
\left[\hbar
v_{F}\left(-\alpha_{1}iD_{x}-\alpha_{2}iD_{y}\right) + \Delta \alpha_{3} + V(\mathbf{r}) - \mathcal{E}  \right]
\Psi( \mathbf{r})=0,
\end{equation}
which describes  low-energy excitations in graphene.
The $4\times 4$ $\alpha$-matrices $\alpha_{i}= \tau_3 \otimes \sigma_i$
and the Pauli matrices $\tau_i$, $\sigma_i$ (as well as the $2\times 2$ unit
matrices $\tau_0$, $\sigma_0$) act on the valley ($\mathbf{K}_\eta $ with
$\eta = \pm$)
and sublattice ($A,B$) indices, respectively, of the
four component spinors $\Psi^T = \left( \Psi_+^T, \Psi_-^T \right) =
\left( \psi_{AK_+}, \psi_{BK_+}, \psi_{BK_-}, \psi_{AK_-} \right)$.
This representation follows  from a tight-binding model for graphene, see e.g. Ref.~\cite{Gusynin2007review} and thus allows one to write down the
appropriate boundary conditions for zigzag and armchair
edges in the continuum model.

We consider both the massless Dirac-Weyl fermions in the pristine graphene
and the massive Dirac fermions with the mass $\Delta/v_F^2$.
Note that a global $A/B$ sublattice asymmetry gap
$2 \Delta \sim \SI{350}{K}$ can be introduced in graphene
\cite{Hunt2013Science,Gorbachev2014Science,Woods2014NatPhys,Chen2014NatCom}
when it is placed on top of hexagonal boron nitride (G/hBN) and
the crystallographic axes of graphene and hBN are aligned.

The  orbital effect of a perpendicular magnetic field $\mathbf{H} = \nabla \times \mathbf{A}$
is included via the covariant spatial
derivative  $D_j=\partial_j+(ie/\hbar c)A_j$ with $j=x,y$ and $-e<0$,
while the potential $V(\mathbf{r})$ corresponds to the
static electric field $e \mathbf{E} =\nabla V(\mathbf{r}) $.
Since moderate values of magnetic field are considered,
the Zeeman energy is small and neglected in this paper
(see, e.g., Ref.~\cite{Gusynin2007review}).

\begin{figure}[h]
\includegraphics[width=.45\textwidth]{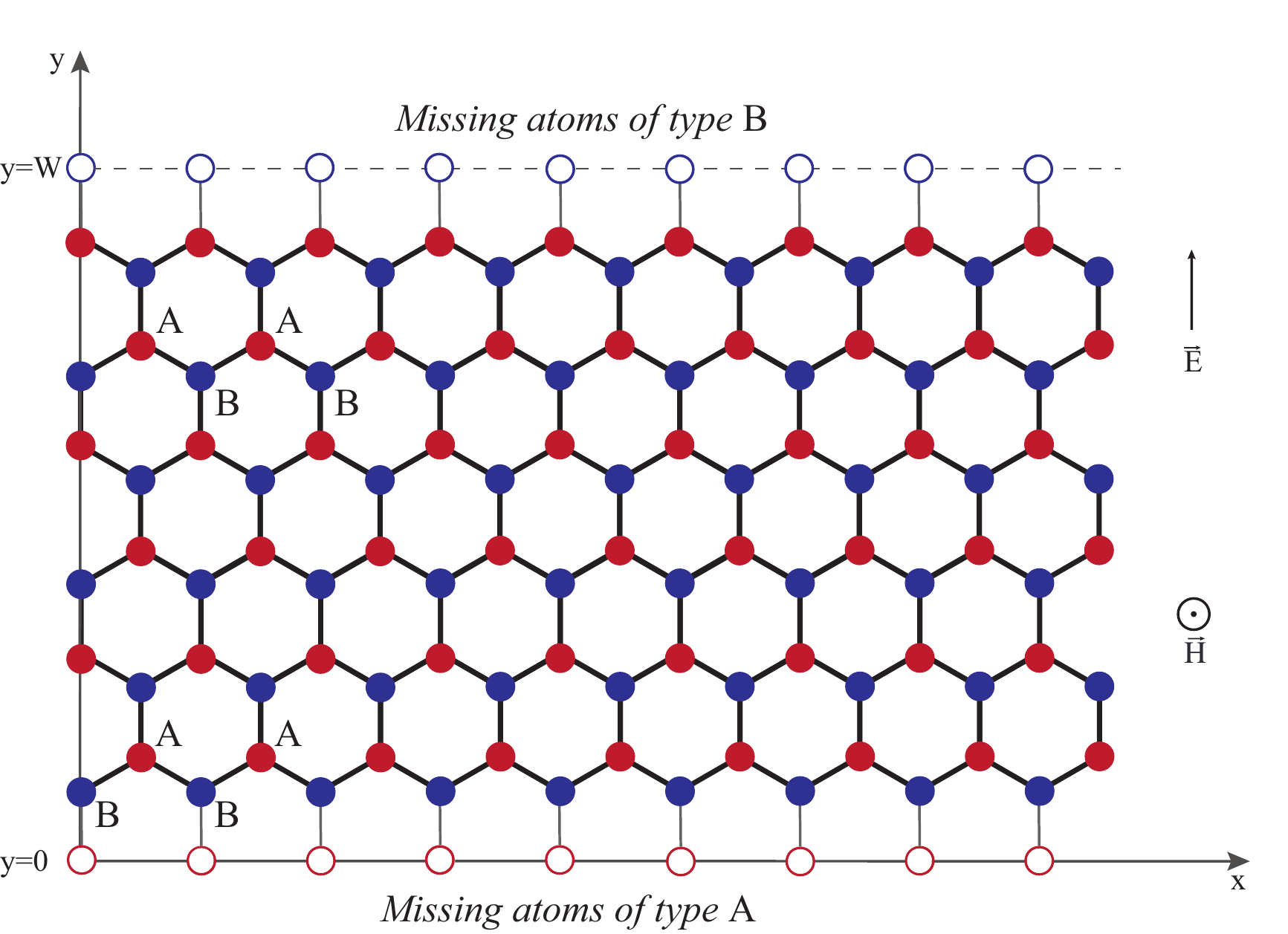}
\caption{
The lattice structure of a finite width graphene ribbon with
zigzag edge and the configuration of applied magnetic and electric fields.}
\label{fig:1}
\end{figure}

We consider the ribbons with the zigzag
as shown in Fig.~\ref{fig:1}.
The ribbons are subjected to the combination
of crossed uniform magnetic and electric fields.
The magnetic field $\mathbf{H}$ applied perpendicular
to the  plane of graphene ribbon along the positive $z$ axis,
and  in-plane  electric field $\mathbf{E}$ applied perpendicular to
the ribbon edges.

%As we will discuss below, only the boundary conditions for the armchair edges
%(\ref{boundary-armchair})
%involve admixture the wavefunctions from the both $\mathbf{K}_{\eta}$ points,
%while
The equation (\ref{Dirac-E}) splits into a pair of two independent
Dirac equations for each $\mathbf{K}_{\eta}$ point:
\begin{equation}
\label{Dirac-eq-2*2}
\left[- i  \hbar v_F \eta (\sigma_1 D_x + \sigma_2 D_y) + \eta \Delta \sigma_3 +
V(\mathbf{r}) -\mathcal{E} \right] \Psi_\eta (\mathbf{r}) =0
\end{equation}
and as we will discuss below, the boundary conditions for
the zigzag edges (\ref{boundary-zigzag}) also do not mix
the corresponding wave functions.

One can see from Eq.~(\ref{Dirac-eq-2*2}) that having the solutions for $\mathbf{K}_+$
point, the corresponding solutions for $\mathbf{K}_-$ can be obtained by
changing the signes of energy $\mathcal{E}$ and electric field in $V(\mathbf{r})$.
Finally, one should take into account that for the spinor $\Psi_-$ the
components of the spinor corresponding to $A$ and $B$ sublattices are
exchanged as compared to $\Psi_+$.

%\subsection{Zigzag edge}
%\label{sec:zigzag}

A zigzag edge is parallel to the $x$ direction as shown in Fig.~\ref{fig:1}.
The in-plane  electric field $\mathbf{E}$ is applied in
the $y$ direction, so the potential $V(\mathbf{r}) = e E y$.
The vector potential is taken in the Landau gauge,
$(A_{x},A_{y})=(-Hy,0)$, where
$H$ is the magnitude of a constant
magnetic field orthogonal to the graphene plane.

Accordingly, the differential equations in
Eq.~(\ref{Dirac-E}) do not depend explicitly on the $x$ coordinate.
Therefore, the wave functions are plane waves in the $x$ direction,
\begin{equation}
\label{x-zigzag}
\begin{split}
\negthickspace \negthickspace \psi_{\scriptscriptstyle{AK_{+}}}(\mathbf{r},k)&=\frac{e^{-ikx}}
{\sqrt{2\pi l}}u_{\scriptscriptstyle{+}}(y,k), \thickspace
\psi_{\scriptscriptstyle{BK_{+}}}(\mathbf{r},k)=\frac{e^{-ikx}}{\sqrt{2\pi
l}}v_{\scriptscriptstyle{+}}(y,k),  \\
\negthickspace \negthickspace
\psi_{\scriptscriptstyle{AK_{-}}}(\mathbf{r},k)&=\frac{e^{-ikx}}
{\sqrt{2\pi l}}u_{\scriptscriptstyle{-}}(y,k), \thickspace
\psi_{\scriptscriptstyle{BK_{-}}}(\mathbf{r},k)=\frac{e^{-ikx}}
{\sqrt{2\pi l}}v_{\scriptscriptstyle{-}}(y,k).
\end{split}
\end{equation}
The wave vector $k$ measures the displacement from $\mathbf{K}_{\eta}$ points.
A particular choice of the coordinate system in Ref.~\cite{Gusynin2007review}
determines that $\mathbf{K}_{\pm} = \pm (2 \pi/a) \, (2/3,0)$,
where $a$ is the lattice constant.
The maximum value of the wave vector $k$
is limited by the boundaries of the first Brillouin
zone.

Recall that the wave vector $k$ determines the center of the
electron orbit along the $y$ direction, $y_0= - kl^2$. Then, as we
shall see below, for a system with a ribbon geometry, e.g.,
$0 \leq y \leq W$, the condition that the peak of the wave function
is inside the ribbon will be satisfied only
for eigenstates with wave vectors $k$ in a finite range, $ - W/l^2 \leq k \leq 0$. This is known as the position -- wave vector duality in
the Landau gauge. Note that the values of the total
wave vector for different $\mathbf{K}_{\eta}$ valleys
in the tight-binding  calculation fall
in different wave vectors domains,
because $K_{+x}\neq K_{-x}$.

Substituting Eq.~(\ref{x-zigzag}) in Eq.~(\ref{Dirac-eq-2*2}) we
obtain the following system of equations for the $\mathbf{K}_{+}$
point
\begin{equation}
\label{zigzag-1}
	\begin{pmatrix}
	\frac{e E y- \mathcal{E} + \Delta}{\hbar v_F} & - \partial_y
-  k - \frac{e}{\hbar c} H y  \\
	 \partial_y -  k -  \frac{e}{\hbar c} H y  & \frac{e E y- \mathcal{E} -\Delta}{\hbar v_F}
	\end{pmatrix} \psi_+ = 0,
	\end{equation}
where $\psi_+^T = (u_+,v_+)$. One can see that
the envelope functions $u_{+}(y,k)$ and $v_{+}(y,k)$
($u_{-}(y,k)$ and $v_{-}(y,k)$) depend
only on a single dimensionless combination of the variables,
$\xi=y/l + kl$ with $l = \sqrt{\hbar c/ |e H|}$ being the magnetic length,
so Eq.~(\ref{zigzag-1}) acquires the form
\begin{equation}
\label{Dirac-zigzag-dimensionless}
	\begin{pmatrix}
	\beta \xi -\epsilon + \delta   &  -i\partial_{\xi} - i \xi \\
	-i\partial_{\xi} + i \xi & \beta \xi-\epsilon - \delta
	\end{pmatrix} \tilde \psi_+ = 0.
\end{equation}
Here we introduced the notations
\begin{equation}
\label{dimensionless}
\beta = \frac{c E}{v_F H}, \quad	
\epsilon =  \frac{l \mathcal{E}}{\hbar v_F} +  l \beta k , \qquad
\delta = \frac{l \Delta}{\hbar v_F} .	
\end{equation}
Writing Eq.~(\ref{Dirac-zigzag-dimensionless}) we used
the spinor ${\tilde \psi}_+^T = (u_+, -i v_+)$. This notation
together with the opposite sign in $\exp (-ikx)$ as compared to
\cite{Gusynin2008PRB,Gusynin2008FNT,Gusynin2009PRB} allows us to
unify the equations describing zigzag and armchair edges.
The latter will be considered in a separate publication.
Here we only remind that the dispersionless edge mode
is absent in the case of  the armchair edge.

The important dimensionless parameter $\beta$ in Eq.~(\ref{dimensionless})
describes the strength of the electric field relative to the
magnetic field. In this paper, we restrict ourselves to
the $|\beta| \leq 1$ case and do not consider pair creation regime.

To obtain  the energy spectrum we need to supplement
the differential equations for the envelope functions $u_{\pm}(y,k)$ and $v_{\pm}(y,k)$ with suitable boundary conditions.
Such conditions can be derived from the tight-binding
model \cite{McCann2004JPCM,Brey2006aPRB,Brey2006bPRB,Ababnin2006PRL}.

In the case of a graphene ribbon of a finite width in the $y$
direction, $0 \leq y \leq W$, and with two zigzag edges parallel to
the $x$ direction, the $A$ and $B$ components of wave functions
should vanish on the opposite edges:
\begin{subequations}
\label{boundary-zigzag}
\begin{align}
\label{boundaryZ-0}
y  & = 0:  \qquad \quad u_{+}(kl)  = u_{-}(kl) =0, \\
\label{boundaryZ-W}
y  & = W:  \qquad v_{+}(W/l+kl)  = v_{-}(W/l+kl) =0,
\end{align}
\end{subequations}
see Fig.~\ref{fig:1}.
Note that the case of the armchair edges is different,
because in the tight-binding calculation
the values of the total wave vector projected
on the armchair edge direction coincide
for the different $\mathbf{K}_{\eta}$ valleys
(see e.g. Ref.~\cite{Deplace2010PRB}). This
allows valley admixing by the boundary condition.

\section{General solutions}
\label{sec:general-sol}

As mentioned in Introduction, the Dirac equation (\ref{Dirac-eq-2*2})
for the massless case, $\Delta =0$ and infinite plane was solved
in Refs.~\cite{Lukose2007PRL,Peres2007JPCM}.
Here we apply a different analytic approach for a finite system.

The main equation (\ref{Dirac-zigzag-dimensionless})
can be rewritten in the following form
\begin{equation}
\label{chi-tilde}
\partial_\xi %\tilde{\chi}(\xi)
{\tilde \psi}_+ (\xi)
=\left(\tilde{A}+ \tilde{B} \xi \right) {\tilde \psi}_+ (\xi)
%\tilde{\chi}(\xi),
\end{equation}
where the $2\times2$  $\xi$-independent matrices $\tilde{A}, \tilde{B}$
are, respectively,
\begin{equation}
\label{tilde-A-B}
\tilde{A}  =
\begin{pmatrix}
0 & i(\epsilon + \delta)\\
i(\epsilon - \delta) & 0
\end{pmatrix}, \qquad
\tilde{B}  =
\begin{pmatrix}
1 & -i\beta \\
-i\beta & -1
\end{pmatrix}.
\end{equation}

Now we make the transformation ${\tilde \psi}_+ =P \chi$ with the matrix $P$ which diagonalizes the matrix $\tilde{B}$, where the matrix
\begin{equation}
P=\begin{pmatrix}
i\gamma & i \\
1 & \gamma
\end{pmatrix},\qquad \gamma=\frac{\beta}{1+\sqrt{1-\beta^2}}.
\label{matrix-P}
\end{equation}
Thus one transforms the system (\ref{chi-tilde}) to the form
\begin{equation}
\label{chi-system}
\partial_\xi \chi(\xi)=\left(A+B \xi \right)\chi(\xi)
\end{equation}
with
\begin{equation}
A=P^{-1} \tilde{A} P =
\begin{pmatrix}
-\frac{\beta \epsilon}{\sqrt{1-\beta^2}} & \delta-\frac{\epsilon}{\sqrt{1-\beta^2}}\\
\delta+\frac{\epsilon}{\sqrt{1-\beta^2}} &
\frac{\beta \epsilon}{\sqrt{1-\beta^2}}
\end{pmatrix},
\end{equation}
and
\begin{equation}
B=P^{-1} \tilde{B} P =
\begin{pmatrix}
-\sqrt{1-\beta^2} & 0 \\
0 & \sqrt{1-\beta^2}
\end{pmatrix} .
\end{equation}
Note that while the present problem in the crossed uniform  fields in
the Cartesian coordinates is exactly solvable by diagonalizing the matrix $B$,
the problem with the radial electric field \cite{Nimyi2022PRB}
which involves three matrices
${\hat A}/\rho+ {\hat B}+{\hat C} \rho$ with $\rho$ being the
radial variable cannot  be solved analytically.

Introducing a new variable $\zeta = (1-\beta^2)^{1/4} \xi+\beta \epsilon / (1-\beta^2)^{3/4}$, we rewrite Eq.~(\ref{chi-system}) as follows
\begin{equation}
\label{chi-system-z}
\partial_\zeta \chi(\zeta)=
\begin{pmatrix}
-\zeta & \sqrt{2} \kappa_{-}\\
\sqrt{2} \kappa_{+} & \zeta
\end{pmatrix}
\chi(\zeta),
\end{equation}
where the following notations are introduced
\begin{equation}
\label{kappa}
\kappa_{\pm}=\frac{\delta \sqrt{1-\beta^2} \pm \epsilon }
{\sqrt{2}(1-\beta^2)^{3/4}}.
\end{equation}

Then one can express the component $\chi_1$ via $\chi_2$,
\begin{equation}
\label{chi1}
\chi_1(\zeta)=\frac{\chi_2'(\zeta)-\zeta \chi_2(\zeta)}{ \sqrt{2} \kappa_{+}},
\end{equation}
and obtain the following equation for $\chi_2$:
\begin{equation}
\label{chi2}
\chi_2''(\zeta)-(1+\zeta^2+ 2 \kappa_{-} \kappa_{+} )\chi_2(\zeta)=0.
\end{equation}
Here prime denotes a differentiation over $\zeta$.

The general solution of the last equation can be written in terms
of the parabolic cylinder (Weber) functions $U(a, x )$ and $V(a, x )$  \cite{Abramowitz.book}
\begin{equation}
\label{chi2-sol}
\chi_2(\zeta) =C_1 U\left(a,\sqrt{2}\zeta \right)
+C_2 V\left(a,\sqrt{2}\zeta \right)
\end{equation}
with
\begin{equation}
\label{a}
a=\frac{1}{2}+  \kappa_{+} \kappa_{-}
=\frac{1}{2}+\frac{\delta^2(1-\beta^2)-\epsilon^2}{2(1-\beta^2)^{3/2}}.
\end{equation}
and $C_{1,2}$ being the integration constants.
Accordingly, using the recurrence relations (19.6.2) and (19.6.5)
from \cite{Abramowitz.book}:
\begin{equation}
\begin{split}
& U'(a,x)-\frac{x}{2}U(a,x)+U(a-1,x)=0,\\
& V'(a,x)-\frac{x}{2}V(a,x)-(a-\frac{1}{2})V(a-1,x)=0
\end{split}
\end{equation}
we obtain from Eq.~(\ref{chi1}) the expression for $\chi_1$
component
\begin{equation}
\label{chi1-sol}
\begin{split}
\chi_1(\zeta)=-\frac{1}{\kappa_{+}} & \Bigg[
C_1 U \left(a-1,\sqrt{2}\zeta \right)  \\
& \left. -C_2\left(a-\frac{1}{2}\right) V\left(a-1,\sqrt{2}\zeta \right) \right].
\end{split}
\end{equation}
One can notice from the definition of $\zeta$ that for a finite
electric field, the electron- and hole-like
solutions become asymmetric.

Finally, one should return to the original spinor components
for the sublattices,  ${\tilde \psi}_+  = P\chi$ (see  Appendix~\ref{sec:Appendix-sol}).
The representation of the solution Eqs.~(\ref{sol-tilde-chi1})
and (\ref{sol-tilde-chi2})
(see also Eqs.~(\ref{chi1-sol}) and (\ref{chi2-sol}))
in terms of the Weber parabolic cylinder functions
$U(a, x )$ and $V(a, x )$ is particulary  convenient because
their Wronskian $W \left\{U,V \right\} = \sqrt{2/\pi}$
\cite{Abramowitz.book} is independent of the parameters.

Another advantage of utilizing these functions is that in
an infinite system without boundaries, the normalizable wave functions
contain only the parabolic cylinder $U(a,x)$ functions.
This is because the cylinder functions $V(a,x)$ diverge exponentially for both positive and negative $x$, as shown by the asymptotics (\ref{V-asymp-x-positive}) and (\ref{V-asymp-x-negative}), which necessitates $C_2$ to be equal to zero.
As one can see from Eq.~(\ref{U-asymp-x-positive}) the remaining $U(a,x)$
functions are bound at $x \to + \infty$. Furthermore, it follows from Eq.~(\ref{V-asymp-x-negative})) that the condition of their finiteness
at $x \to -\infty$ requires that $a=-n-1/2$ with $n$ being a non-negative
integer, viz.
\begin{equation}
\label{LL-collapse-dimensionless}
\epsilon_n= (1-\beta^2)^{1/2}
\begin{cases}
- \eta \delta \, \mbox{sgn}(eH), & n=0, \\
\pm\sqrt{2n(1-\beta^2)^{1/2}+\delta^2},& n=1,2\ldots
\end{cases}
\end{equation}
Thus we recover the result of Refs.~\cite{Lukose2007PRL,Peres2007JPCM}
generalized for a finite $\Delta$ case (see Ref.~\cite{Alisultanov2014PB}
and recent works \cite{Arjona20017PRB,Nimyi2022PRB}):
\begin{equation}
\label{LL-collapse}
\begin{split}
\mathcal{E}_n & = \mathcal{E}^\ast_n - \hbar k \frac{c E}{H}, \\
\mathcal{E}^\ast_n & = (1 - \beta^2 )^{1/2} \\
\times & \begin{cases}
- \eta \Delta,  & n=0, \\
\pm \sqrt{\frac{ 2n \hbar v_F^2 e H}{c}
(1 - \beta^2 )^{1/2} + \Delta^2 }, & n=1,2\ldots
\end{cases}
\end{split}
\end{equation}
Here and in what follows we assume that $H > 0$.
Obviously the collapse of Landau levels occurs as $|\beta|$
approaches $1$.

\section{Zigzag edge states for a ribbon}
\label{sec:ribbom-zigzag}

To comprehend a link between the existing computations on the
lattice and the considered here continuum model we begin
with numerical solution of the equations
(\ref{spectrum-gen-zigzag}) and (\ref{spectrum-gen-zigzag-K-})
derived in Appendix~\ref{sec:Appendix-sol}.
They determine dimensionless energies
$\epsilon_\alpha=\epsilon_n(kl,W/l)$ as functions of quantum numbers
$\alpha\equiv (n, k)$ and the ratio $W/l$
for the $\mathbf{K}_+$ and $\mathbf{K}_-$ valleys, respectively.
To return to the total energy $\mathcal{E}$ (or its dimensionless analog
$l \mathcal{E}/(\hbar v_F)$) one should use
Eq.~(\ref{dimensionless}) to restore its linear in $k$
part.

\subsection{The gapless  case, $\Delta=0$}
\label{sec:zigzag-gapless-ribbon}

\begin{figure}[!ht]
\includegraphics[width=.48\textwidth]{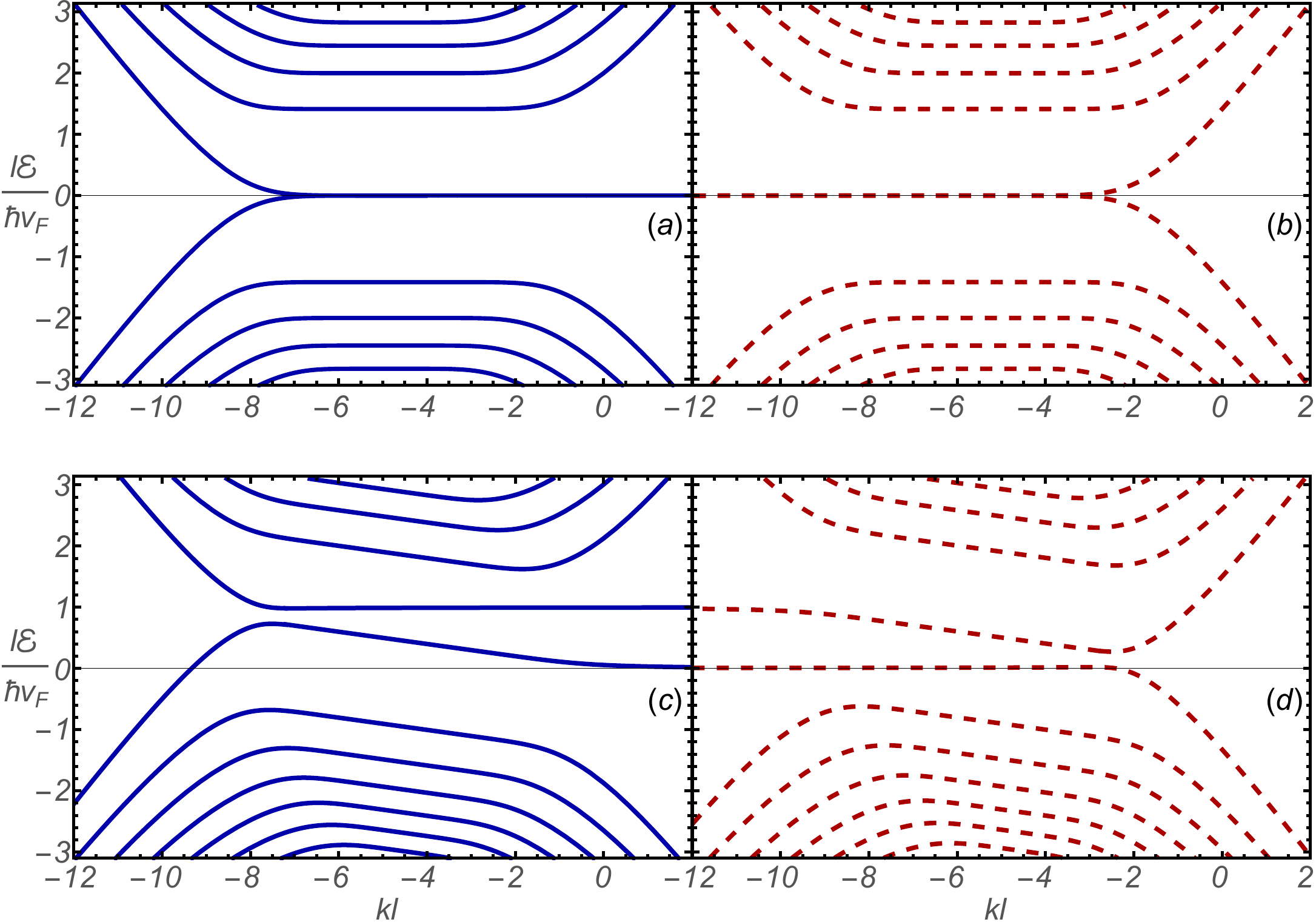}
\caption{The energy spectra $l \mathcal{E} (k)/(\hbar v_F)$
of the first few Landau levels of the ribbon of widths $W = 10l$
%near a zigzag edge of graphene
for the gapless, $\Delta=0$ case.
Panels (a) and (b) show the results for $\beta =0$ and
(c) and (d) for $\beta =0.1$, respectively.
The solutions for the $\mathbf{K}_+$ valley are shown
in panels (a) and (c) by the solid (blue) lines,
the solutions for the $\mathbf{K}_-$ valley
in panels (b) and (d) by the dashed (red) lines.
}
\label{fig:zigzag-ribbon-gapless}
\end{figure}

The corresponding spectra for $\Delta =0$ case are computed numerically and
presented  in Fig.~\ref{fig:zigzag-ribbon-gapless}.
The panels~(a), (b) show the results for zero and the panels~(c), (d)
for a finite ($\beta =0.1$) electric field.
Since the wave vector $k$ is counted from $K_{\pm x}$
the spectra for both valleys fell on the same range of
the wave vectors as compared
to the solutions of the lattice model
\cite{Lukose2007PRL,Roslyak2010PTRS,Ostahie2015PRB},
where the wave vector includes the value $K_{\pm x}$.
In our case the edges correspond to the
values $k l =-10$  and $k l =0$ for both valleys,
but to facilitate the comparison we showed
the results for the $\mathbf{K}_+$ and $\mathbf{K}_-$ valleys
side by side.
%Considering that the wave vector $k$ has the opposite sign,
One finds that the results presented in Fig.~\ref{fig:zigzag-ribbon-gapless}
are in agreement with the calculations done for the
lattice model in a finite electric field
\cite{Lukose2007PRL,Ostahie2015PRB} (notice the opposite sign of $k$).

In Fig.~\ref{fig:zigzag-ribbon-gapless}~(a), (b)
we observe the dispersionless Landau levels in the
bulk of the ribbon. The $n \neq 0$ Landau levels are dispersing
independently near the edges for each valley, while for
the lowest $n=0$ Landau level there is a dispersionless part
of the spectrum branch that unites both valleys together.
These dispersionless surface states localized at the boundaries
\cite{Fujita1996JPSJ,Brey2006aPRB,Brey2006bPRB,Ababnin2006PRL}
and $n=0$ Landau level form the degenerate states.

On the contrary,
Fig.~\ref{fig:zigzag-ribbon-gapless}~(c), (d)
show that the degeneracy is lifted by an applied
electric field \cite{Lukose2007PRL}. All Landau levels including
the lowest one develop a linear $k$ dispersion with the slope
proportional to the electric field, whereas only
the surface states remain dispersionless.

\subsection{The gapped  case}
\label{sec:zigzag-gapped-ribbon}

\begin{figure}[!ht]
\includegraphics[width=.48\textwidth]{{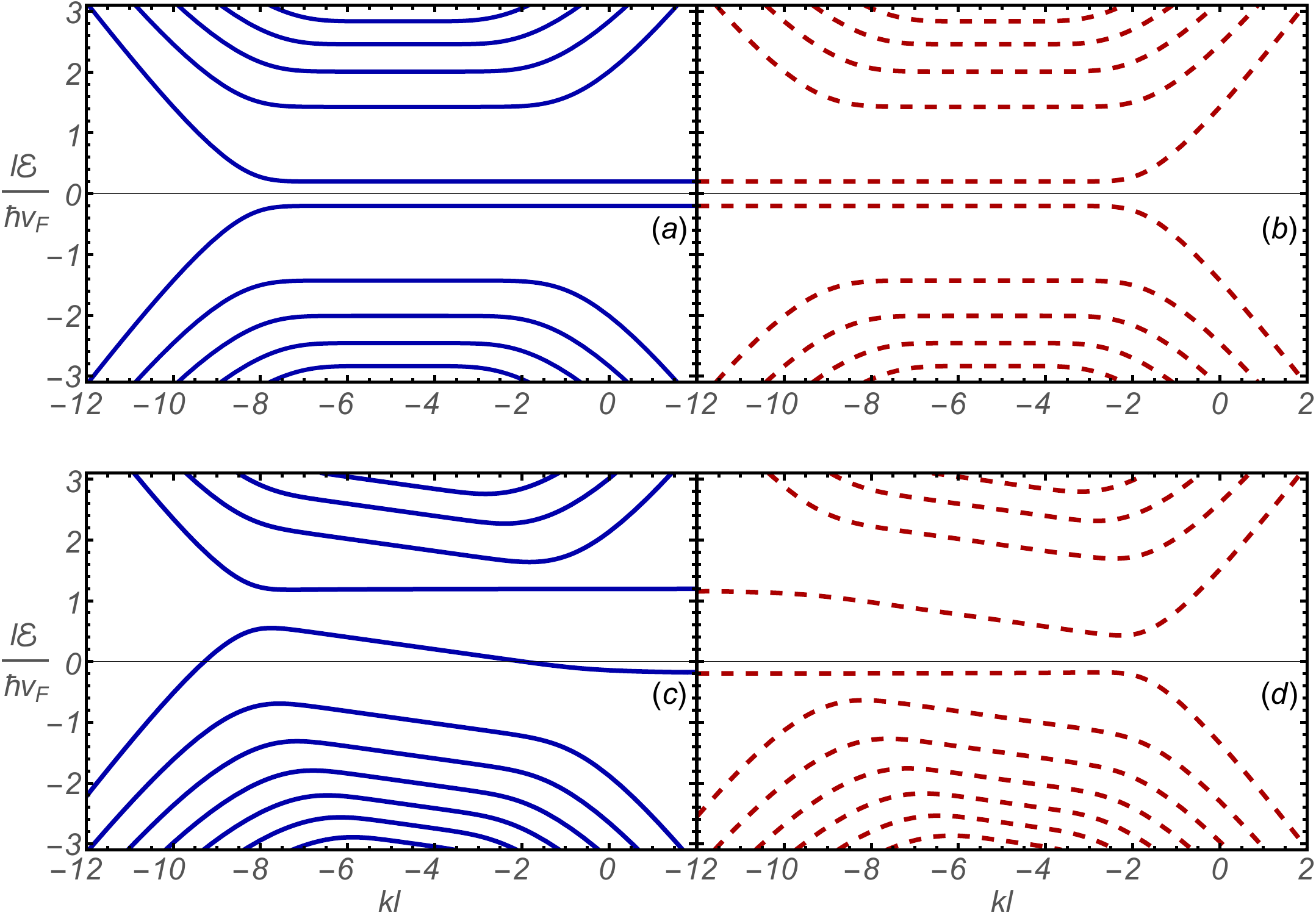}}
\caption{The energy spectra $l \mathcal{E} (k)/(\hbar v_F)$
of the first few Landau levels of the ribbon of widths $W = 10 l$
for the gapped, $\delta= 0.2$ case.
The panels (a), (b), (c) and (d) are for the same valleys
and values of $\beta$ as in  Fig.~\ref{fig:zigzag-ribbon-gapless}.
}
\label{fig:zigzag-ribbon-gapped}
\end{figure}

The corresponding spectra for a finite $\Delta$ case are
computed numerically and presented  in Fig.~\ref{fig:zigzag-ribbon-gapped}.
The corresponding numerical solution shown in Figs.~\ref{fig:zigzag-ribbon-gapped}~(a) and (b) is in agreement with the results presented in Refs.~\cite{Gusynin2009PRB} for zero electric field.
The gap $\Delta$ considered in the present work corresponds to
the parity breaking and the time reversal symmetry conserving gap $\tilde \Delta$
in the notation of \cite{Gusynin2008PRB,Gusynin2008FNT,Gusynin2009PRB}.

The presence of the mass gap results in the absence of the gapless edge
states, the degeneracy of the $n=0$ level is lifted.
The degeneracy of  the surface modes is also lifted,
yet they remain dispersionless with the energies $\sim \pm \Delta$.

In the presence of an electric field
[see Figs.~\ref{fig:zigzag-ribbon-gapped}~(c) and (d)]
all Landau levels acquire linear $k$ dispersion,
but the surface modes remain dispersionless.
Furthermore, while for the $\Delta =0$ case
the linearly dispersing part of the $n=0$ level
has the same energy, for a finite $\Delta$ this
degeneracy is lifted as for the $\beta =0$ case.

To study the discussed features by analytic methods we
need to simplify the problem by considering the  semi-infinite geometry.

\section{Zigzag edge in semi-infinite geometry}
\label{sec:zigzag-semi}

On a half-plane, normalizable wave functions are also given in terms of only
$U(a, x)$ function which as mentioned above
falls off exponentially as $x \to \infty$, while the function $V(a, z )$
is growing exponentially in both directions $x \to \pm \infty$.
Therefore, we must take again $C_2 =0$. Yet,
in contrast to the case of an infinite
plane, on a half-plane, there is no restriction for the parameter
$a$ to be a negative half-integer [see the discussion above Eq.~(\ref{LL-collapse-dimensionless})].

The zigzag boundary conditions (\ref{boundaryZ-W})
at $y=W \to \infty$ are automatically satisfied due to the
asymptotic (\ref{U-asymp-x-positive}).
It follows from the remaining boundary conditions
(\ref{boundaryZ-0}) at $y=0$ that
the term with $C_1$ in Eq.~(\ref{sol-tilde-chi1})
has to be zero. The latter condition produces the following equation
for the spectrum for the $\mathbf{K}_+$ valley:
\begin{equation}
\label{spectrum-zigzag+half}
\gamma U \left(a-1, \sqrt{2} \zeta (0) \right)
- \kappa_{+} U\left(a, \sqrt{2} \zeta (0) \right) =0.
\end{equation}
%Here $\gamma$  and $\kappa_{+}$   are defined by Eqs.~(\ref{matrix-P})
%and (\ref{kappa}), respectively.
Using the asymptotics
Eq.~(\ref{asymp-x-positive}) one can verify that
Eq.~(\ref{spectrum-zigzag+half}) also follows from
Eq.~(\ref{spectrum-gen-zigzag}) in the limit $W \to \infty$.

To write down the corresponding equation
for the spectrum for the $\mathbf{K}_-$ valley we use the prescriptions
described below Eq.~(\ref{Dirac-eq-2*2}). They imply that
$\epsilon \to - \epsilon$, $\beta \to - \beta$
and $\psi_1 \leftrightarrow  \psi_2$, so that
we arrive at the following equation
\begin{equation}
\label{spectrum-zigzag-half}
U \left(a-1, \sqrt{2} \zeta (0) \right)
+ \gamma \kappa_{-} U\left(a, \sqrt{2} \zeta (0) \right) =0.
\end{equation}
Eqs.~(\ref{spectrum-zigzag+half}) and (\ref{spectrum-zigzag-half})
determine dimensionless energies $\epsilon_\alpha=\epsilon_n(kl)$
as functions of quantum numbers $\alpha\equiv (n, k)$.
To return to the total energy $\mathcal{E}$ (or its dimensionless analog
$l \mathcal{E}/(\hbar v_F)$) one should use Eq.~(\ref{dimensionless})
to restore its linear in $k$ part.

\begin{figure}[!ht]
\includegraphics[width=.42\textwidth]{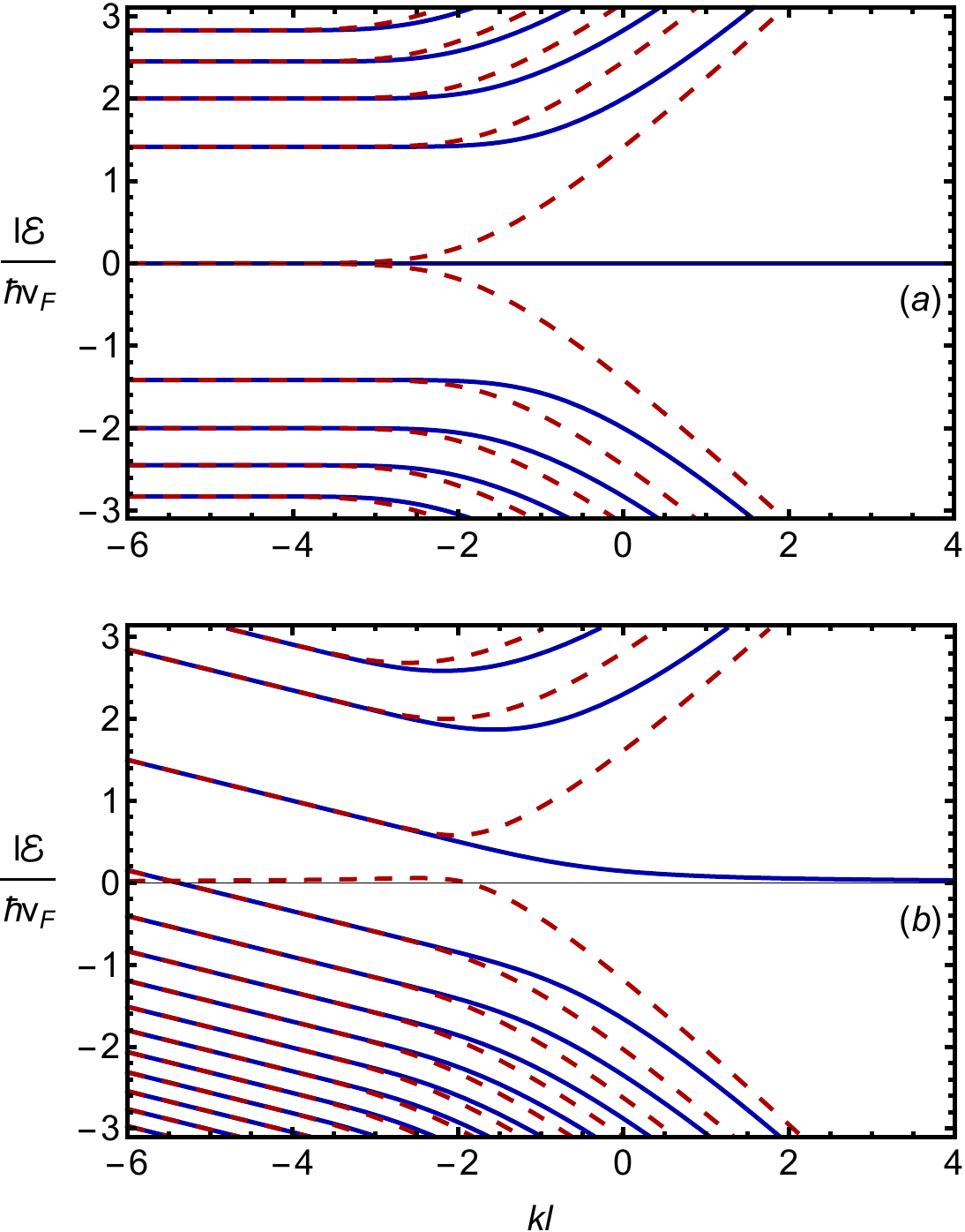}
\caption{The energy spectra $l \mathcal{E} (k)/(\hbar v_F)$
of the first few Landau levels near a zigzag edge of graphene
for the gapless, $\Delta=0$ case. The solutions for the
$\mathbf{K}_+$ and $\mathbf{K}_-$ valleys are shown by the
solid (blue) and dashed (red) lines, respectively.
(a) $\beta =0$; (b) $\beta =0.25$
}
\label{fig:zigzag-spectrum}
\end{figure}

\begin{figure}[!ht]
\includegraphics[width=.42\textwidth]{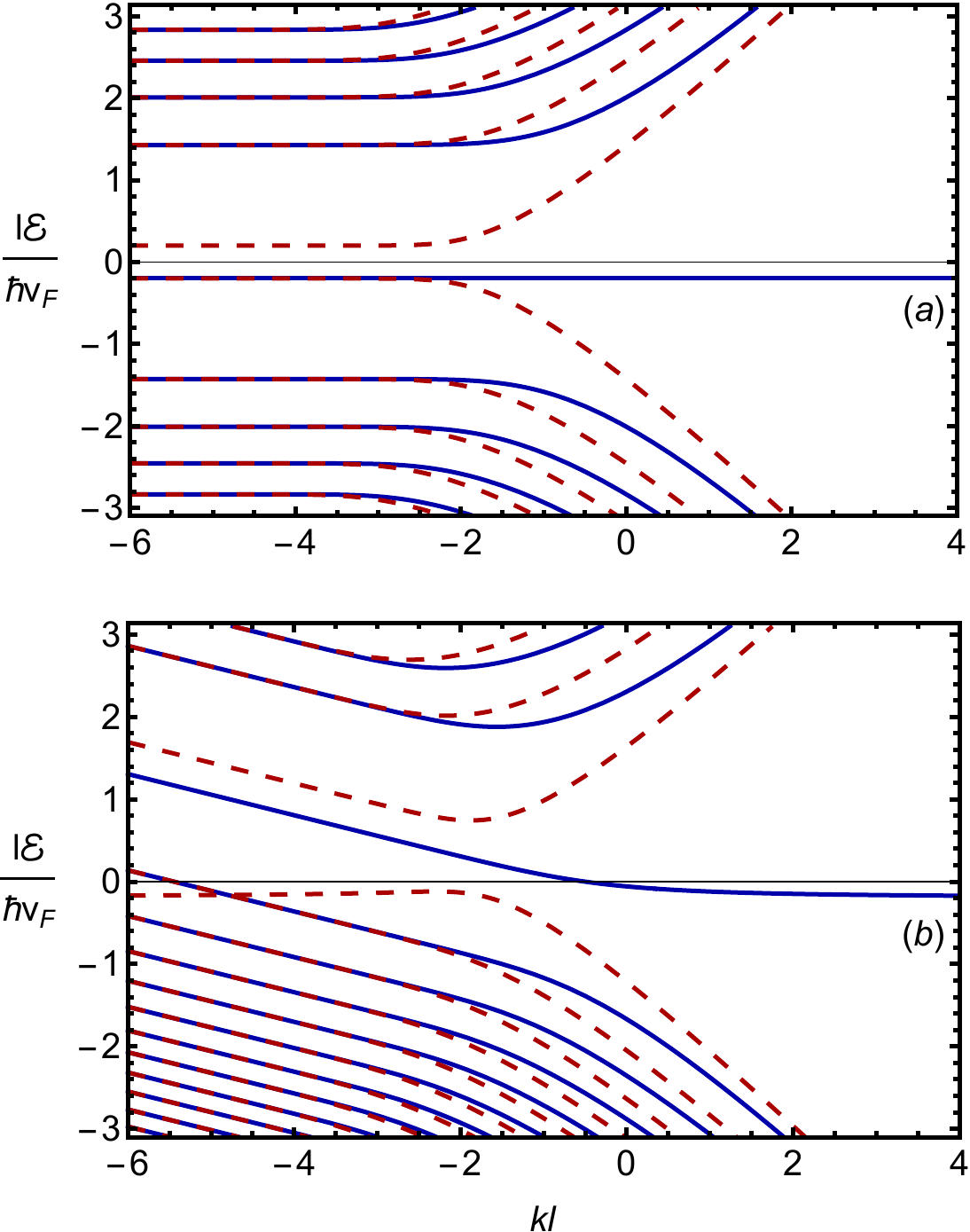}
\caption{The energy spectra $l \mathcal{E} (k)/(\hbar v_F)$
of the first few Landau levels near a zigzag edge of graphene
for the gapped, $\delta=0.2$ case.
The panels (a) and (b) are for the same
values of $\beta$ as in  Fig.~\ref{fig:zigzag-spectrum}.
}
\label{fig:zigzag-gapped-spectrum}
\end{figure}

The corresponding spectra are computed numerically and
presented  for the gapless and gapped cases in
Figs.~\ref{fig:zigzag-spectrum} and \ref{fig:zigzag-gapped-spectrum},
respectively.
In contrast to
Figs.~\ref{fig:zigzag-ribbon-gapless} and \ref{fig:zigzag-ribbon-gapped}
where we showed the $\mathbf{K}_{+}$ and $\mathbf{K}_{-}$ valleys on separate
panels, here we superimpose both valleys on the same
panel to allow a direct comparison of the
corresponding energy levels. This is possible, because
in the continuum model  the wave vector $k$ is counted from $K_{\pm x}$
values. The negative values of $k$ correspond to the bulk,
while the edge is at $k=0$.

Since a half-plane geometry is considered, for a finite
$\beta$ there is an unbound linear growth of the dispersion curves as
$kl \to -\infty$. The presence of the other edge at $y=W$
modifies this behavior. The hole-like levels including the
blue line (the $\mathbf{K}_+$ solution) that goes to zero
for $kl \to \infty$ would go downward, while the
electron-like levels go upward. Furthermore,
the degeneracy of the solutions for the $\mathbf{K}_+$
and $\mathbf{K}_-$ valleys would be lifted near the other edge
as one can see in Figs.~\ref{fig:zigzag-ribbon-gapless} and \ref{fig:zigzag-ribbon-gapped}.

The only curve  which remains dispersionless as $kl \to -\infty$
even in the ribbon geometry is the zero energy lower branch of
the $\mathbf{K}_-$ valley spectrum
that corresponds to the surface states mentioned in Sec.~\ref{sec:zigzag-gapless-ribbon}.
The second edge of the ribbon supports a second dispersionless mode
which is absent in a half-plane geometry. The surface state
in the semi-infinite geometry will be discussed in
Sec.~\ref{sec:dispersionless}.

\subsection{Zero electric field limit}

In the absence of electric field, $\beta = \gamma =0$ and
$a = [1 + \delta^2 - \epsilon^2]/2$, so
Eqs.~(\ref{spectrum-zigzag+half}) and (\ref{spectrum-zigzag-half})
for the spectra in the $\mathbf{K}_\pm$ valleys
are in agreement with the equations studied in Refs.~\cite{Gusynin2008PRB,Gusynin2008FNT}
\begin{equation}
\label{E=0-eqs}
(\delta + \epsilon) U \left(a,\sqrt{2} kl \right) =0, \qquad
U \left(a-1, \sqrt{2} kl \right) =0.
\end{equation}
The corresponding numerical solutions shown in Figs.~\ref{fig:zigzag-spectrum}~(a)
and \ref{fig:zigzag-gapped-spectrum}~(a)
are in agreement with the results presented in Refs.~\cite{Ababnin2006PRL,Brey2006bPRB,Gusynin2008PRB,Gusynin2008FNT}.

It is known that the zigzag edge  hosts a band of dispersionless zero-energy states localized at the edge even in the absence of magnetic
and electric fields \cite{Fujita1996JPSJ,Brey2006aPRB,Brey2006bPRB,Ababnin2006PRL}.

In the presence of a magnetic field, the lowest $n=0$ Landau
level coexists with this surface state and behaves
rather differently for the two valleys.
As discussed in  \cite{Ababnin2006PRL,Romanosvsky2011PRB},
the lowest Landau level for the $\mathbf{K}_{+}$ valley  near
the edge transforms into the surface mode [see the blue straight line
in Fig.~\ref{fig:zigzag-spectrum}~(a)].
For the $\mathbf{K}_{-}$ valley the lowest Landau level and the
surface state mix, producing two dispersing edge modes as can be seen from the behavior of the red dashed lines in Fig.~\ref{fig:zigzag-spectrum}~(a).

It is possible to find out the approximate solutions of Eqs.~(\ref{E=0-eqs})
in the vicinity of the $y=0$ edge and for the bulk. The derivation
follows along the lines of the corresponding derivation
for the Shr\"{o}dinger equation \cite{Heuser1974ZP,Yerin2021PRB}.
First of all one can verify that the dispersionless
solution $\mathcal{E}_0 = - \Delta$
of the first equation in Eqs.~(\ref{E=0-eqs}) for the $\mathbf{K}_{+}$ valley
indeed satisfies the original system (\ref{chi-system-z}).

Depending on the relationship between
the $y$ coordinate of the center of the
electron cyclotron orbit, $y_0= - kl^2$,
and the magnetic length $l$ one can obtain two types of
the asymptotic solutions of Eqs.~(\ref{E=0-eqs}).

For electrons near the boundary, $y_0 \ll l$,
Eqs.~(\ref{E=0-eqs}) are solved by expanding the function $U(a,z)$
in the argument $z$ (see Eqs.~(19.3.5) in \cite{Abramowitz.book}).
In the first order in $z$ we obtain the following  equation
$\Gamma\left(1/4+a/2\right) \Gamma^{-1}\left(3/4+a/2\right)=\sqrt{2}z$
for the $\mathbf{K}_+$ valley, while for the $\mathbf{K}_-$ valley $a \to a-1$.
Exactly at the edge for $z = -\sqrt{2} y_0 =0$
the eigenenergies are determined by the poles of the gamma function
$\Gamma(z)$ in the denominator at $z = -n$ with $n=0,1,2, \ldots$
These eigenvalues do not depend of $y_0$.
Then we use the expansion
$\Gamma^{-1} (-n - \alpha) \simeq (-1)^n \Gamma(n+1) (-\alpha)$
in order to obtain correction $\alpha$  to these eigenenergies
dependent on $y_0$.

For the bulk electrons, $y_0 \gg l$, the argument of the parabolic
functions is large and negative, so that one uses firstly
the relationship (\ref{U(-x)}) which brings in the gamma
function $\Gamma(1/2+a)$ in the denominator.
Then  after using the asymptotic expressions (\ref{asymp-x-positive})
essentially the same procedure as in the previous case is applied.
As the result we obtain the following expressions for the
$\mathbf{K}_+$ valley
\begin{equation}
\label{sol-K+}
\begin{split}
& \epsilon_{+,n}^2 - \delta^2 =
\begin{cases}
2n + \frac{2^n}{\sqrt{\pi}(n-1)!}
\left(\frac{y_0}{l} \right)^{2n-1} e^{-y_0^2/l^2} , & y_0 \gg l, \\
4n \left[1- \frac{\Gamma(n+1/2)}{\pi n!}\frac{2 y_0}{l} \right]
,& y_0 \ll l,
\end{cases}
\end{split}
\end{equation}
with $n =1,2, \ldots$ and for the $\mathbf{K}_-$ valley
\begin{equation}
\label{sol-K-}
\begin{split}
& \epsilon_{-,n}^2 - \delta^2 =
\begin{cases}
2n + \frac{2^{n+1}}{\sqrt{\pi}n!}
\left(\frac{y_0}{l} \right)^{2n+1}
e^{- y_0^2/l^2} , & y_0 \gg l, \\
2(2n+1) \left[1- \frac{\Gamma(n+1/2)}{\pi n!}\frac{2 y_0}{l} \right]
,& y_0 \ll l,
\end{cases}
\end{split}
\end{equation}
with $n =0,1,2, \ldots$
The level $\epsilon_{-,0} \approx - \delta$ is related to the surface
mode and it disappears when the edge is moved to
infinity, similarly to the disappearance of another
surface mode when one goes from the ribbon to a
half-plane geometry.
Thus the combined spectrum for both $\mathbf{K}_{\pm}$ valleys reduces
the one to given by Eq.~(\ref{LL-collapse-dimensionless}) with
$\beta =0$.

By studying the spectra of the edge and bulk regions,
we can observe an interesting property: for the
$\mathbf{K}_+$ valley, we have $\epsilon_{+,n}(y_0=0) = \epsilon_{+,2n}(y_0 \gg l)$ with $n>0$, and for the $\mathbf{K}_-$ valley, we have $\epsilon_{-,n}(y_0=0) = \epsilon_{-,2n+1}(y_0 \gg l)$ with $n \geq 0$ \cite{Deplace2010PRB,Romanosvsky2011PRB}.
This feature of the spectra can be traced back to the
character of the spectrum for the harmonic oscillator with
reflecting wall at the minimum of potential (see the Problem
2.5 in Ref.~\cite{Galitski.book}). The difference
between the valleys is caused by the difference of the first
arguments of $U$ in Eqs.~(\ref{E=0-eqs}). When the opposite
edge is also considered, these properties of the spectra
for the $\mathbf{K}_+$ and $\mathbf{K}_-$ valleys are interchanged.

\subsection{The lowest Landau level and surface mode in a finite electric field}
\label{sec:dispersionless}

As discussed at the end of Sec.~\ref{sec:zigzag-gapless-ribbon},
the electric field lifts the degeneracy of
the $n=0$ Landau level and the dispersionless state.
Indeed, in Fig.~\ref{fig:zigzag-spectrum}~(b) one observes
splitting of the two red (dashed) curves for the $\mathbf{K}_-$ valley that
merge to zero energy in Fig.~\ref{fig:zigzag-spectrum}~(a)
as $kl \to -\infty$. The upper curve corresponds to the
dispersing $n=0$ level, while the lower curve
is related to the dispersionless surface state.
As we saw in the case of the ribbon shown Fig.~\ref{fig:zigzag-ribbon-gapless}~(c) and (d) in the $\mathbf{K}_+$ valley this state evolves
in the dispersing lowest $n=0$  Landau level whose energy
decreases as $kl \to -\infty$. In a half-plane geometry the
corresponding blue curve increases linearly as $kl \to -\infty$.

It is interesting to investigate the origin of
the dispersionless state. Indeed, in our approach
all branches of the spectrum are obtained from Eq.~(\ref{dimensionless}),
viz. $l \mathcal{E}/(\hbar v_F) = \epsilon - \beta k l$.
This implies that the linear dispersion of the
Landau levels in the bulk is related to the $-\beta l k$
($-\hbar k E/H$ in the dimensional units) term,
while the dispersionless mode has to emerge from a delicate
cancelation with the $\epsilon$ term that should
show the same dependence on $kl$.
By using the expansion of parabolic cylinder functions of large order and argument as developed by Darwin \cite{Abramowitz.book}, Appendix~\ref{sec:Appendix-dispersionless} demonstrates that a solution exists for the mode where $\epsilon \sim \beta l k$ as $k l \to -\infty$.

Furthermore, one can show that there is also
a similar surface mode solution in the $\mathbf{K}_+$
valley  which shows the same behavior as $k l \to \infty$.

For the $\Delta =0$ and finite $\beta$ the dispersive parts
of the curves corresponding to the $n=0$ level
for the $\mathbf{K}_+$  and $\mathbf{K}_-$ valleys have the same
energy [see Fig.~\ref{fig:zigzag-spectrum}~(b)].
As shown in Fig.~\ref{fig:zigzag-gapped-spectrum}~(b),
for a finite $\Delta$ this degeneracy is lifted as for the $\beta =0$ case.
The dispersionless mode in the gapped case tends
to $\sim - \Delta$ value.

\subsection{Landau levels with $n \neq 0$ in a finite electric field}

The behaviour of the $n \neq 0$ Landau levels is similar for
$\beta =0$ and finite $\beta$, viz.
there are two branches of the edge states, one for each
valley. These states are degenerate in the bulk and split near
the edge. An increase in the electric field deforms the spectrum in two ways: firstly, the familiar linear term $\hbar k E/H$, which is present in nonrelativistic systems, appears; and secondly, the energy levels
themselves merge as $|\beta|$ approaches $1$. Both these effects are
indeed observed for the computations done for the larger values of
$\beta$ than shown in Figs.~\ref{fig:zigzag-spectrum}~(b)
and \ref{fig:zigzag-gapped-spectrum}~(b).
The second feature reflects the Landau level collapse.

\section{Conclusion}
\label{sec:conclusion}

We conducted both analytical and numerical studies on the edge states of a graphene ribbon with finite width and semi-infinite geometry utilizing low-energy theory
in the presence of the crossed magnetic and electric fields. Our findings are consistent with numerical calculations performed on a lattice. When an electric field is present, all Landau levels exhibit linear wave vector dispersion, except for surface modes that remain dispersionless. We devoted special attention to the analytical study of these states and demonstrated that they can be obtained using Darwin's expansion of the parabolic cylinder functions of large order and argument.

The existence of states localized near the zigzag edges of a graphene ribbon in the absence of a magnetic field was confirmed by STS measurements \cite{Niimi2006PRL}. More recently, the signatures of these states were observed in magnetotransport measurements \cite{Wu2018PRL}. Extending these measurements to cases where an electric field is applied to the ribbon would be useful.

\begin{acknowledgments}
We would like to thank the Armed Forces of Ukraine for providing security to perform this work.
The authors acknowledge a support by the National Research Foundation of Ukraine grant  (2020.02/0051) ``Topological phases of matter and excitations in Dirac materials, Josephson junctions and magnets''.
We would like to thank O.V.~Bugaiko
for participating in this work at its initial stage.
We would like to express our gratitude to
I.A.~Shovkovy for insightful discussions on numerical methods.

\end{acknowledgments}

\appendix

\section{Explicit form of the solution and equations
for eigenenergies}
\label{sec:Appendix-sol}

The explicit form of the general solution for
the components of the spinor $\tilde \chi$ reads
\begin{equation}
\label{sol-tilde-chi1}
\begin{aligned}
%\tilde \chi_1(\zeta)
{\tilde \psi}_{+1} (\zeta)
&=i C_{+1}
\left[ \gamma U\left(a-1,\sqrt{2}\zeta \right)-\kappa_{+}
U\left(a,\sqrt{2}\zeta \right)\right]\\
+  i  C_{+2} & \left[ \gamma
V\left(a-1,\sqrt{2}\zeta\right)+\frac{\kappa_{+}}{a-1/2}
V \left(a,\sqrt{2}\zeta \right)\right],
\end{aligned}
\end{equation}
\begin{equation}
\label{sol-tilde-chi2}
\begin{aligned}
%\tilde  \chi_2(\zeta)
{\tilde \psi}_{+2} (\zeta)
&= C_{+1} \left[  U\left(a-1,\sqrt{2}\zeta \right)-
\gamma \kappa_{+}
U\left(a,\sqrt{2}\zeta \right)\right]\\
+  C_{+2} & \left[  V \left(a-1,\sqrt{2}\zeta \right)+
\frac{\gamma \kappa_{+}}{a-1/2}V \left(a,\sqrt{2}\zeta \right)\right],
\end{aligned}
\end{equation}
where for the convenience of further analysis we redefined the integration
constants $C_{1,2}$ in Eqs.~(\ref{chi1-sol})
and (\ref{chi2-sol}).
Writing Eqs.~(\ref{sol-tilde-chi1}) and (\ref{sol-tilde-chi2})
we used $\kappa_{+}$, $a$ and $\gamma$ defined by
Eqs.~(\ref{kappa}), (\ref{a}) and (\ref{matrix-P}), respectively.

The asymptotic behavior of the functions $U(a,x)$ and $V(a,x)$ for the
large positive $x$ is the following
(see Eqs.~(19.8.1) and (19.8.2) from \cite{Abramowitz.book}):
\begin{subequations}
\label{asymp-x-positive}
\begin{align}
\label{U-asymp-x-positive}
&U(a,x\to\infty)\simeq e^{-\frac{x^2}{4}}x^{-a-\frac{1}{2}}
\left[1 + O \left(\frac{1}{x^2} \right) \right],\\
%\qquad x\to\infty,\\
\label{V-asymp-x-positive}
&V(a,x\to\infty)\simeq \sqrt{\frac{2}{\pi}}e^{\frac{x^2}{4}}x^{a-\frac{1}{2}}
\left[1 + O\left(\frac{1}{x^2} \right) \right].
%\qquad x\to\infty.
\end{align}
\end{subequations}

To derive asymptotics of the functions $U(a,x)$, $V(a,x)$ for large negative
$x$ we use Eqs.~(19.4.2), (19.4.3) from \cite{Abramowitz.book}
rewritten as follows
\begin{subequations}
\label{cylindric_functions_relations}
\begin{align}
\label{U(-x)}
&U(a,-x)=\frac{\pi}{\Gamma\left(\frac{1}{2}+a\right)}V(a,x)-\sin\pi a U(a,x),\\
\label{V(-x)}
&V(a,-x)=\sin\pi a V(a,x)+\frac{1}{\pi}\Gamma\left(\frac{1}{2}+a\right)\cos^2\pi a U(a,x)
\end{align}
\end{subequations}
and obtain as $x\to \infty$:
\begin{subequations}
\label{asymp-x-negative}
\begin{align}
\label{U-asymp-x-negative}
&U(a,-x)\simeq\frac{\sqrt{2\pi}}{\Gamma\left(\frac{1}{2}+a\right)}
e^{\frac{x^2}{4}}x^{a-\frac{1}{2}}
\left[1 + O \left(\frac{1}{x^2} \right) \right] , \\
\label{V-asymp-x-negative}
&V(a,-x)\simeq \sqrt{\frac{2}{\pi}}\sin (\pi a )\, e^{\frac{x^2}{4}}x^{a-\frac{1}{2}} \left[1 + O \left(\frac{1}{x^2} \right) \right]. \end{align}
\end{subequations}

It follows from the zigzag boundary conditions
(\ref{boundary-zigzag}) for the $\mathbf{K}_+$ valley that the spectrum
is determined by the following transcendental secular equation
\begin{equation}
\label{spectrum-gen-zigzag}
\begin{split}
& \left[ U \left(a-1, \sqrt{2} \zeta (W) \right) -
\kappa_+ \gamma
U \left(a, \sqrt{2} \zeta (W) \right) \right] \\
\times & \left[
\gamma V \left(a-1,\sqrt{2} \zeta(0) \right) +
\frac{\kappa_+}{a-1/2} V (a, \sqrt{2} \zeta(0) )
 \right] \\
= & \left[ V \left(a-1, \sqrt{2} \zeta (W) \right) +
\frac{\kappa_+ \gamma}{a-1/2}
V \left(a, \sqrt{2} \zeta (W) \right) \right] \\
\times & \left[
\gamma U \left(a-1, \sqrt{2} \zeta (0) \right)
- \kappa_+  U\left(a, \sqrt{2} \zeta (0) \right)\right],
\end{split}
\end{equation}
where
\begin{equation}
\label{zeta-0-W}
\begin{split}
\zeta (0)\equiv & \zeta(y=0)  =  (1-\beta^2)^{1/4} k l
+ \frac{\beta \epsilon} { (1-\beta^2)^{3/4}}, \\
\zeta (W)\equiv & \zeta(y=W)  =
(1-\beta^2)^{1/4} \left(\frac{W}{l} + k l \right) \\
& + \frac{\beta \epsilon} { (1-\beta^2)^{3/4}}.
\end{split}
\end{equation}
Recall also that $\kappa_{\pm}$ and $a$ depend on $\epsilon$.
Using the prescriptions described above Eq.~(\ref{spectrum-zigzag-half})
one can write the corresponding secular equation for the spectrum
for the $\mathbf{K}_-$ valley:
\begin{equation}
\label{spectrum-gen-zigzag-K-}
\begin{split}
&\left[ U\left( a-1,\sqrt{2} \zeta (0)\right) +\gamma \kappa_- U
\left(a,\sqrt{2} \zeta(0) \right)\right]\\
\times& \left[\gamma V\left( a-1,\sqrt{2} \zeta (W)\right)
-\frac{\kappa_-}{a-\frac{1}{2}} U \left(a,\sqrt{2} \zeta(0)
\right)\right]\\
=& \left[\gamma U\left( a-1,\sqrt{2} \zeta (W)\right) + \kappa_- U
\left(a,\sqrt{2} \zeta(W) \right)\right]\\
\times& \left[ V\left( a-1,\sqrt{2} \zeta (0)\right) -\frac{\gamma
\kappa_-}{a-\frac{1}{2}} V \left(a,\sqrt{2} \zeta(0) \right)\right].
\end{split}
\end{equation}

\section{Solution for the dispersionless mode at the $\mathbf{K}_-$ valley}
\label{sec:Appendix-dispersionless}

We seek for a linear $\epsilon\simeq Akl$ solution of
Eq.~(\ref{spectrum-zigzag-half}) in the limit
$kl\to-\infty$. It is convenient
to rewrite this equation introducing the following notations
$\mathfrak{a} \equiv a -1/2 = \kappa_{+}\kappa_{-}$ and
$x \equiv - \sqrt{2} \zeta (0) \to\infty$:
\begin{equation}
\label{dispersionless-eq-K-}
U(\mathfrak{a}-1/2,-x)+\gamma\kappa_{-}U(\mathfrak{a}+1/2,-x) =0.
\end{equation}
Similarly to the derivation of the
asymptotic solutions (\ref{sol-K+}) and (\ref{sol-K-}) for the
$y_0 \gg l$ case when the argument of the parabolic
functions is large and negative, one uses firstly
the relationship (\ref{U(-x)}) written in the following
form
\begin{equation}
U(a,-x)=\Gamma(1/2-a)\cos\pi a V(a,x)-\sin\pi a U(a,x).
\end{equation}
However, in contract to the abovementioned case
of Eqs.~(\ref{sol-K+}) and (\ref{sol-K-}) for $y_0 \gg l$,
one cannot rely on the asymptotic expannsions
(\ref{asymp-x-positive}), because the argument $a$ of
the parabolic functions also goes to $-\infty$.

Thus one has to use another expansion of the parabolic
functions applicable for $a<0$, $x\to+\infty$
and $x^2+4a \to \infty$. This case corresponds to the
Darwin's expansion given by Eqs.~(19.10.6) and (19.10.7)
from \cite{Abramowitz.book}:
\begin{subequations}
\begin{align}
&U(a,x)=\frac{\sqrt{\Gamma(1/2-a)}}{(2\pi)^{1/4}}\exp\left[-\theta(a,x)+v(a,x)\right],\\
&V(a,x)=\frac{2}{(2\pi)^{1/4}\sqrt{\Gamma(1/2-a)}}\exp\left[\theta(a,x)+v(a,-x)\right],
\end{align}
\end{subequations}
where
\begin{align}
& \theta(a,x)=\frac{1}{4}x X(a)+a\ln\frac{x+X(a)}{2\sqrt{|a|}}, \\
& v(a,x)\sim -\frac{1}{2}\ln X(a)+\sum_{s=1}(-1)^s\frac{d_{3s}}{X^{3s}(a)},
\end{align}
with $X(a)=\sqrt{x^2-4|a|}$,
and the coefficients $d_{3s}$ are given in Eq.~(19.10.13) in \cite{Abramowitz.book}. We are interested in the case $X(a)\gg1$,
so the terms with $d_{3s}$ can be neglected.
Furthermore, the functions $U (\mathfrak{a}\pm1/2,x)$ that contain the
decaying  exponents
$\exp [-\theta(\mathfrak{a}\pm1/2,x)]$ in the $x\to \infty$ limit
may also be neglected, so that we are left with the following asymptotic expressions
\begin{equation}
\label{U-Darwin}
\begin{split}
&U(\mathfrak{a}- 1/2,-x)=\\
&\frac{2\sin\pi \mathfrak{a} \sqrt{\Gamma(1-\mathfrak{a})}}{(2\pi)^{1/4}}
X^{-1/2} (\mathfrak{a}-1/2,x)  e^{\theta(\mathfrak{a}-1/2,x)},\\
&U(\mathfrak{a}+1/2,-x)= \\
&-\frac{2\sin\pi \mathfrak{a} \sqrt{\Gamma(-\mathfrak{a})}}{(2\pi)^{1/4}}
X^{-1/2}(\mathfrak{a}+1/2,x)
e^{\theta(\mathfrak{a}+1/2,x)}
\end{split}
\end{equation}
that has to be substituted into Eq.~(\ref{dispersionless-eq-K-}).
For $|a| \gg 1$ the functions $X^{-1/2} (\mathfrak{a} + \lambda,x)$ and
$\theta(\mathfrak{a}+ \lambda,x)$ with $\lambda = \pm 1/2$ can be expanded
as follows
\begin{equation}
\label{X-theta-expansions}
\begin{split}
& X^{-1/2}(\mathfrak{a}+\lambda)=X^{-1/2}(\mathfrak{a})-
\frac{\lambda}{X^{5/2}(\mathfrak{a})},\\
& \theta(\mathfrak{a}+\lambda)=
\theta(\mathfrak{a})+\lambda\ln\frac{x+X(\mathfrak{a})}{2\sqrt{-a}}
+O\left(\frac{x}{\mathfrak{a}X(\mathfrak{a})}\right).
\end{split}
\end{equation}

To rewrite the final equation that relates $\epsilon$ and $kl$
in the considered limit in the most elucidating form,
we introduce the new notations
$\mathfrak{a} = -\mu^2/2$ and $x=\mu t\sqrt{2}$.
Then using Eq.~(\ref{a}) and the value $\zeta(0)$ given by
Eq.~(\ref{zeta-0-W}) we obtain that
\begin{equation}
\label{mu-t-defintions}
\mu=\frac{|\epsilon|}{(1-\beta^2)^{3/4}},\qquad t=(1-\beta^2)\frac{kl}{\varepsilon}+\beta.
\end{equation}
Then one can see that
\begin{equation}
\label{Temme-function}
\begin{split}
& \ln\frac{x+X(\mathfrak{a})}{2\sqrt{- \mathfrak{a}}}=\ln(t+\sqrt{t^2-1}),\\
&  \theta(\mathfrak{a},x)=\frac{\mu^2}{2}\left[t\sqrt{t^2-1}+\ln(t+\sqrt{t^2-1}\right].
\end{split}
\end{equation}
Note also that these notations also allow to establish a
link between the used here Darwin's expansion \cite{Abramowitz.book}
and the expansions of the Weber parabolic cylinder functions
obtained in \cite{Temme2000} (see also \cite{NIST-book}).

Substituting the representations (\ref{U-Darwin}) in Eq.~(\ref{dispersionless-eq-K-}), where
$X^{-1/2} (\mathfrak{a} + \lambda,x)$ and
$\theta(\mathfrak{a}+ \lambda,x)$ given by  the expansions (\ref{X-theta-expansions})
are rewritten using Eq.~(\ref{Temme-function}), we arrive
at very simple equation for unknown $t$:
\begin{equation}
\sqrt{- \mathfrak{a} }=\gamma\kappa_{-}(t+\sqrt{t^2-1}).
\end{equation}
Taking into account that for $\delta =0$ the parameter
$\kappa_{-}=\mu/\sqrt{2}$  the last equation reduces to
\begin{equation}
\gamma(t+\sqrt{t^2-1})=1,
\end{equation}
which has a solution $t=1/\beta$. Therefore,
using the definition (\ref{mu-t-defintions}) of $t$
we find that $\varepsilon=\beta kl$ for $k l \to -\infty$.

\end{document}